\documentclass[lettersize,journal]{IEEEtran}
\usepackage{amsmath,amsfonts}
\usepackage{algorithmic}
\usepackage{algorithm}
\usepackage{array}
\usepackage[caption=false,font=normalsize,labelfont=sf,textfont=sf]{subfig}
\usepackage{textcomp}
\usepackage{stfloats}
\usepackage{url}
\usepackage{verbatim}
\usepackage{graphicx}
\usepackage{cite}
\usepackage[table]{xcolor}
\usepackage{makecell}
\usepackage{tabularx}
\usepackage{textcomp}
\usepackage{array}
\usepackage{multirow}

\definecolor{hllightblue}{RGB}{232,242,252}
\newcommand{\tw}{\cellcolor{hllightblue}}
\newcommand{\uw}{$\mu$W}

\newcolumntype{Y}[1]{>{\hsize=#1\hsize\centering\arraybackslash}X}
\newcolumntype{G}[1]{>{\hsize=#1\hsize\columncolor{hllightblue}\centering\arraybackslash}X}
\newcolumntype{M}[1]{>{\centering\arraybackslash}m{#1}}

\usepackage{eso-pic}
\usepackage[hidelinks]{hyperref}
\newcommand{\arxivpublicationnotice}{%
\AddToShipoutPictureFG*{%
\AtPageUpperLeft{\put(\LenToUnit{0.65in},\LenToUnit{-9pt}){\makebox(0,0)[lt]{%
\parbox[t]{7.2in}{\fontsize{7.5}{8.5}\selectfont\raggedright
Author's accepted manuscript. Accepted for publication in \textit{IEEE Solid-State Circuits Letters}, 2026.\\
B. Liu, Y. Hou, Y. Huang, J. Xu, H. You, A. Hung, N. Yang, J. Ma, L. Kondrataviciute, S. K. Kalia, A. G. Richardson, R. Genov, and X. Liu, ``A Neuroprosthetic SoC Featuring Rapid Local Sensory Feedback with On-Chip Unsupervised Multi-Class Spike Sorting.''\\
DOI \href{https://doi.org/10.1109/LSSC.2026.3746977}{10.1109/LSSC.2026.3746977}. IEEE Xplore \url{https://ieeexplore.ieee.org/document/11753970}.
}}}}%
\AtPageLowerLeft{\put(\LenToUnit{0.65in},\LenToUnit{9pt}){\makebox(0,0)[lb]{%
\parbox[b]{7.2in}{\fontsize{7}{8}\selectfont\raggedright
\copyright\ 2026 IEEE. Permission from IEEE must be obtained for all other uses, in any current or future media, including reprinting/republishing this material for advertising or promotional purposes, creating new collective works, for resale or redistribution to servers or lists, or reuse of any copyrighted component of this work in other works.
}}}}%
}}

\begin{document}

\title{A Neuroprosthetic SoC Featuring Rapid Local Sensory Feedback with On-Chip Unsupervised Multi-Class Spike Sorting}

\author{Bowen Liu,
        Yuhan Hou,
        Yu Huang,
        Jianxiong~Xu,
        Hao~You,
        Ashley~Hung,
        Naize~Yang,
        Junyu~Ma,
        Laura~Kondrataviciute,
        Suneil~K.~Kalia,
        Andrew~G.~Richardson,
        Roman~Genov,
        and~Xilin~Liu
\thanks{B. Liu, Y. Hou, Y. Huang, H. You, J. Xu, A. Hung, N. Yang, J. Ma, R. Genov, and X. Liu are with the Edward S. Rogers Sr. Department of Electrical and Computer Engineering, University of Toronto, Toronto, ON M5S 2E4, Canada.}
\thanks{L. Kondrataviciute and S. K. Kalia are with Toronto Western Hospital, Toronto, ON M5T 1M8, Canada.}
\thanks{A. G. Richardson is with the Department of Neurosurgery, University of Pennsylvania, Philadelphia, PA 19104, USA.}    }

\markboth{IEEE Solid-State Circuits Letters}{Liu \MakeLowercase{\textit{et al.}}: Neuroprosthetic SoC}

\maketitle
\thispagestyle{empty}
\arxivpublicationnotice

\begin{abstract}
This letter presents a 65 nm neuroprosthetic system-on-chip (SoC) featuring a rapid feedback pathway from motor recording to sensory stimulation. To enable low-latency sensory mapping modulated by individual-neuron firing rates, the SoC integrates a mixed-signal unsupervised spike-sorting module with stimulation-artifact rejection, a programmable sliding-window firing-rate calculator, and a closed-loop controller for adapting stimulation parameters. The spike-sorting module combines event-driven analog feature extraction with Manhattan-distance-based digital clustering while consuming 1.6\,$\mu$W/channel. On the Quiroga Difficult-1 benchmark, it achieves 94.58\% three-class accuracy, with only a 0.4\% reduction when stimulation artifacts are introduced at 20\% of the spike count. The closed-loop controller supports polynomial and logarithmic mapping functions and incorporates Mitchell's approximation for power saving. The SoC is fully characterized on the bench and \textit{in vivo}.
\end{abstract}


\vspace{-3mm}

\section{Introduction}

Tens of millions worldwide have motor impairments caused by limb loss, paralysis, or neurodegenerative disease~\cite{yu2025neuroprosthetic}. Neuroprostheses translate neural recordings into commands for assistive devices, enabling intention-driven control.
Advances in electrode resolution and recording electronics provide large volumes of data for motor decoders, but gains from additional channels are limited by steep learning curves and inadequate feedback.
Conventional prostheses rely solely on visual feedback, while sensor-driven stimulation adds artificial touch for dexterous manipulation but remains limited by signal-chain latency and trial-and-error learning.
Rapid local feedback between the motor and sensory cortices provides a promising alternative approach~\cite{prsa2017rapid}, in which sensory feedback inferred directly from motor signals modulates sensory stimulation, thereby reducing latency and simplifying learning.
Animal studies using optogenetic stimulation show that rapid feedback can accelerate motor target acquisition~\cite{goueytes2022learning}.
Given the limited clinical translatability of optogenetics, electrical stimulation offers an alternative but requires reliable artifact rejection.

\begin{figure}[!t]
    \centering
    \includegraphics[width=.95\linewidth]{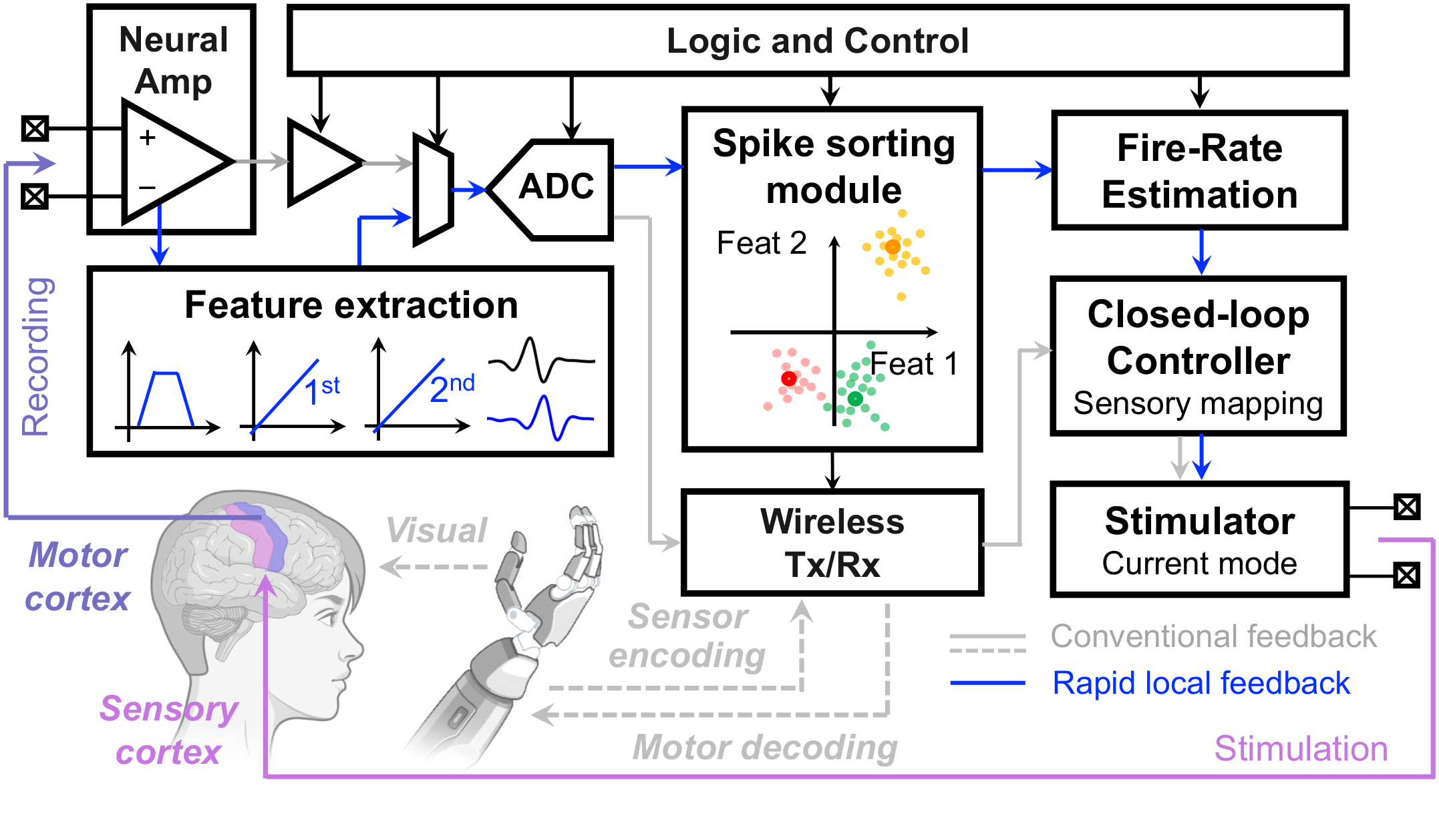} \vspace{-4mm} \vspace{-5pt}
    \caption{Conceptual illustration of the developed neuroprosthetic SoC supporting rapid local sensory modulation in addition to conventional external feedback loops. Key building blocks and signal paths are highlighted.}
    \label{fig:intro} \vspace{-3mm}
\end{figure}

In this work, we present a system-on-chip (SoC) for this emerging class of neuroprostheses with rapid sensory feedback. Fig.~\ref{fig:intro} illustrates its key building blocks and signal paths. The design was initially presented at the International Solid-State Circuits Conference (ISSCC)~\cite{yu2025neuroprosthetic}. This letter expands on the discussion of the SoC’s neuromodulation subsystem, focusing on unsupervised spike sorting, rapid sensory modulation, neural interface design, and experimental validation. The wireless power and data transfer subsystems will be described in a separate manuscript currently in preparation.

The key contributions are as follows. First, an ultralow-power unsupervised Manhattan-distance clustering engine enables accurate, adaptive multiclass spike sorting with robust stimulation-artifact rejection. Second, an energy-efficient neural front-end combines an open-loop (OL) low-noise amplifier with event-driven feature extraction. Third, on-chip closed-loop controllers enable rapid sensory feedback by mapping firing rates of selected neurons over a programmable sliding window to stimulation parameters, using hardware-efficient logarithmic approximation and compact R-2R DACs for current-mode stimulation.

\vspace{-2mm}
\section{System and Circuits Implementation}

\subsection{Unsupervised Adaptive Multi-Class Spike Sorting}

\begin{figure}[!t]
    \centering
    \includegraphics[width=.95\linewidth]{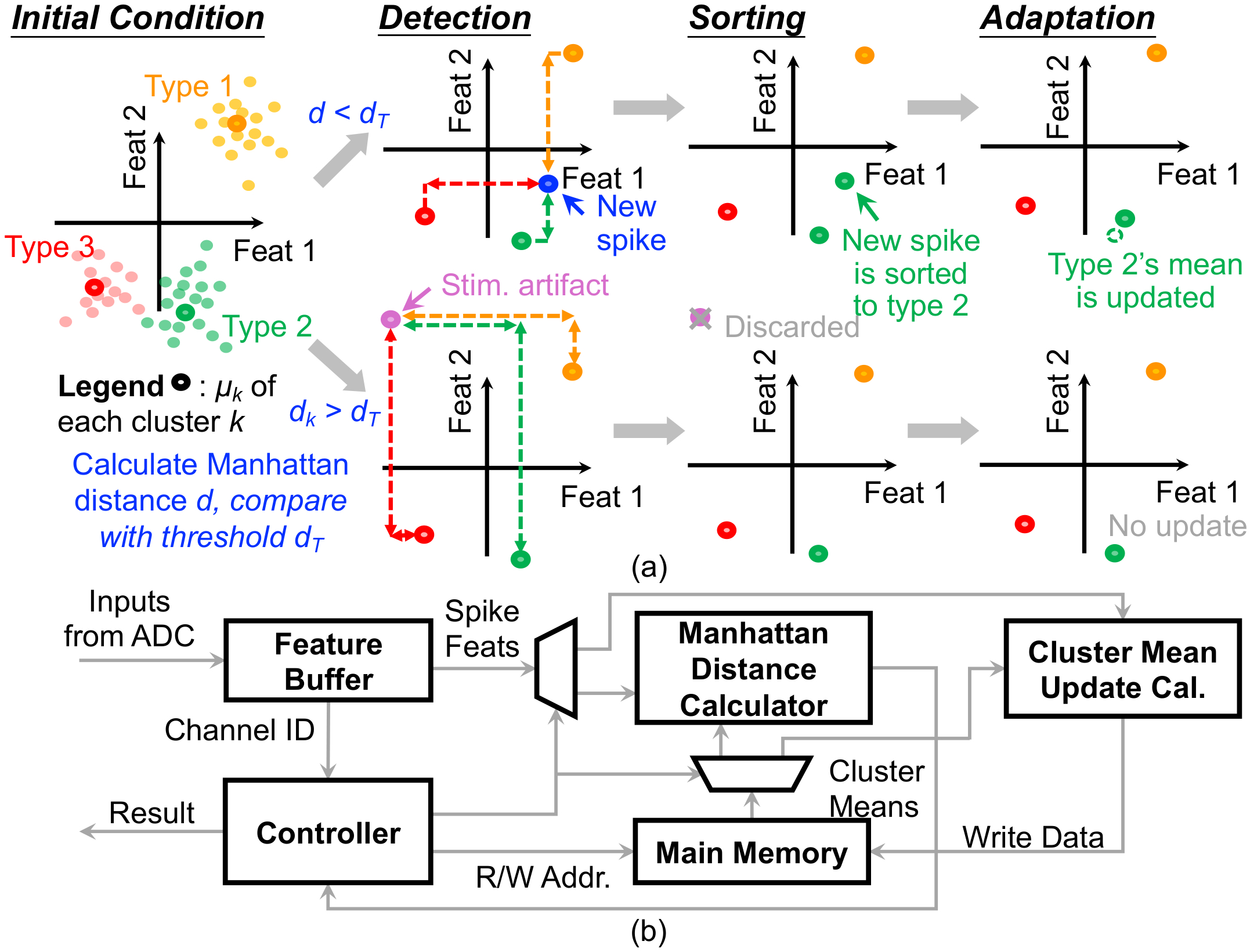} \vspace{-3mm} \vspace{-3pt}
    \caption{(a) Conceptual illustration of this unsupervised adaptive spike sorting process. (b) Simplified block diagram of the Manhattan-distance clustering engine for multi-class spike sorting.} \vspace{-3mm}
    \label{fig:ckt_spike_sorting}\vspace{-1mm}
\end{figure}

The spike sorter uses ordered maximum depolarization and repolarization slopes $f_1$ and $f_2$ to isolate neuron-specific motor activity. Each spike sorter serves 16 channels, with four sorters integrated on chip. Each channel supports a programmable number of clusters, denoted by $k$, while a buffered event-driven engine is shared across all 16 channels to handle coincident events.
Cluster centroids are initialized using K-means++ without labeled data.
For a feature vector $\mathbf{f}=[f_1,f_2]$ and a centroid $\boldsymbol{\mu}_k=[\mu_{k,1},\mu_{k,2}]$, the sorter evaluates the Manhattan distance $d_k=\lvert f_1-\mu_{k,1}\rvert+\lvert f_2-\mu_{k,2}\rvert$ and selects the minimum valid distance (Fig.~\ref{fig:ckt_spike_sorting}(a)). Manhattan distance preserves sufficient cluster separability while avoiding multipliers and square-root operations, substantially reducing sorting power, area, and latency. A threshold $d_{T}$ rejects outliers, such as stimulation and motion artifacts. The controller addresses the corresponding memory entry and streams only that channel's active centroids (Fig.~\ref{fig:ckt_spike_sorting}(b)). It retains the minimum-distance candidate while examining the remaining clusters, then updates the selected centroid after a valid assignment with an update rate of $\eta$.
Dynamic cluster creation, deletion, and merging are not supported for simplicity.
The cluster count $k$ is manually configured but could be estimated automatically in future work.

\begin{figure}[!b]
    \centering \vspace{-6mm}
    \includegraphics[width=.9\linewidth]{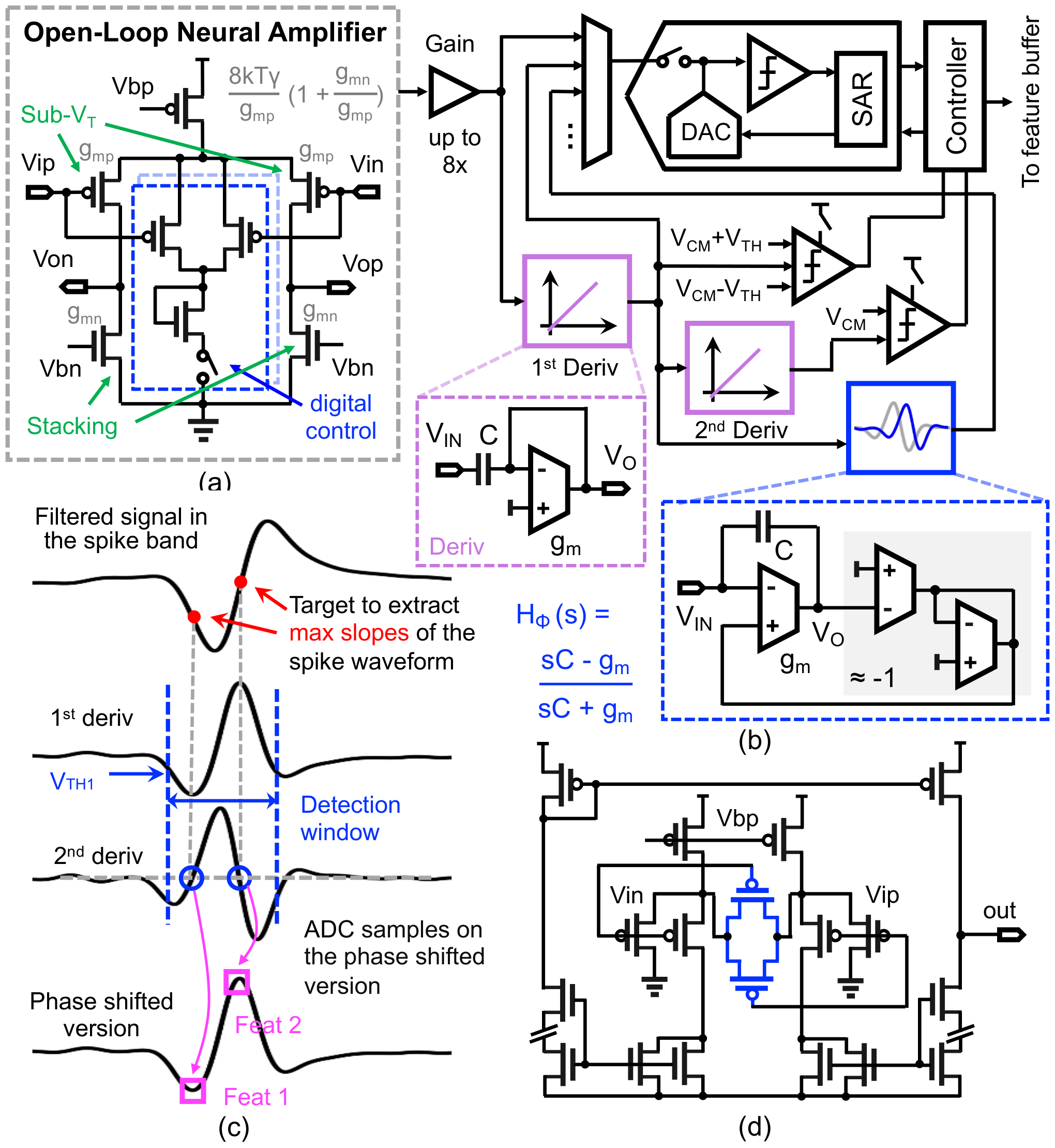} \vspace{-3mm} \vspace{-1pt}
    \caption{(a) Open-loop neural amplifier. (b) Simplified block diagram of the feature extraction circuit including the phase shifting block. (c) Illustration of the feature extraction process. (d) Schematic of the $g_m$ cell.} \label{fig:ckt_analog}
\end{figure}

\vspace{-1mm}
\subsection{Neural Front-End with Event-Driven Feature Extraction}

The recording front-end features a configurable OL neural amplifier and a feature-extraction path (Fig.~\ref{fig:ckt_analog}(a, b)). The OL architecture eliminates feedback components, reducing area and power while still meeting the noise and linearity requirements for spike sorting. The feature extractor finds both slope extrema (Fig.~\ref{fig:ckt_analog}(c)).
A first-derivative threshold $V_{TH}$ (1/16 of full scale by default) opens a detection window, and second-derivative zero crossings trigger ADC conversions at the extrema.
The temporal order is preserved.
To compensate for the detection delay and align the sampled features with the slope extrema, an analog phase shifter is used. Implemented using programmable $g_m$ cells (Fig.~\ref{fig:ckt_analog}(d)), its transfer function is $H_{\phi}(s)=(sC-g_m)/(sC+g_m)$, which has unity magnitude and phase $\phi(\omega)=\pi-2\arctan(\omega C/g_m)$. This first-order Pad\'{e} approximation of $e^{-sT}$ provides a low-frequency group delay of $T=2C/g_m$, enabling tunable delays of 1 to 10\,$\mu$s. For a nominal delay of 1~$\mu$s, simulations across process variations and temperatures from 30 to 50~$^\circ$C show worst-case delay deviations of +65/-54~ns before calibration. The phase shifter introduces a worst-case timing jitter of 1.453~ns, which is negligible relative to the timescale of spike signals.

\vspace{-3mm}
\subsection{Firing-Rate Calculation and Closed-loop Mapping}

\begin{figure}[!t]
    \centering \includegraphics[width=.98\linewidth]{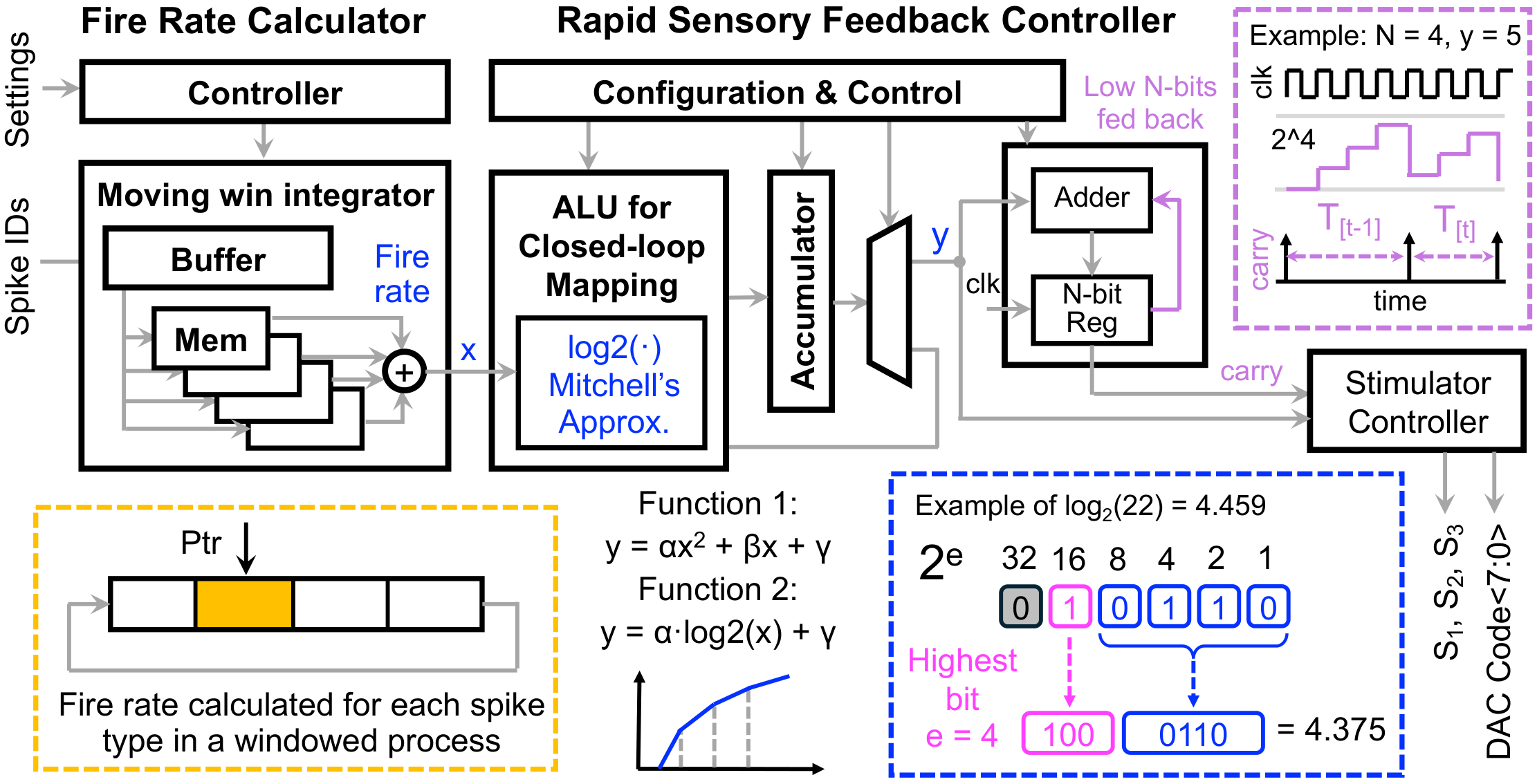} \vspace{-3mm} \vspace{-1pt}
    \caption{Moving-window firing-rate calculator and programmable sensory-feedback controller with Mitchell's approximation and stimulation mapping.} \vspace{-3mm}
    \label{fig:ckt_cl_controller}
\end{figure}

\begin{figure}[!b]
    \centering \vspace{-4mm}
    \includegraphics[width=.98\linewidth]{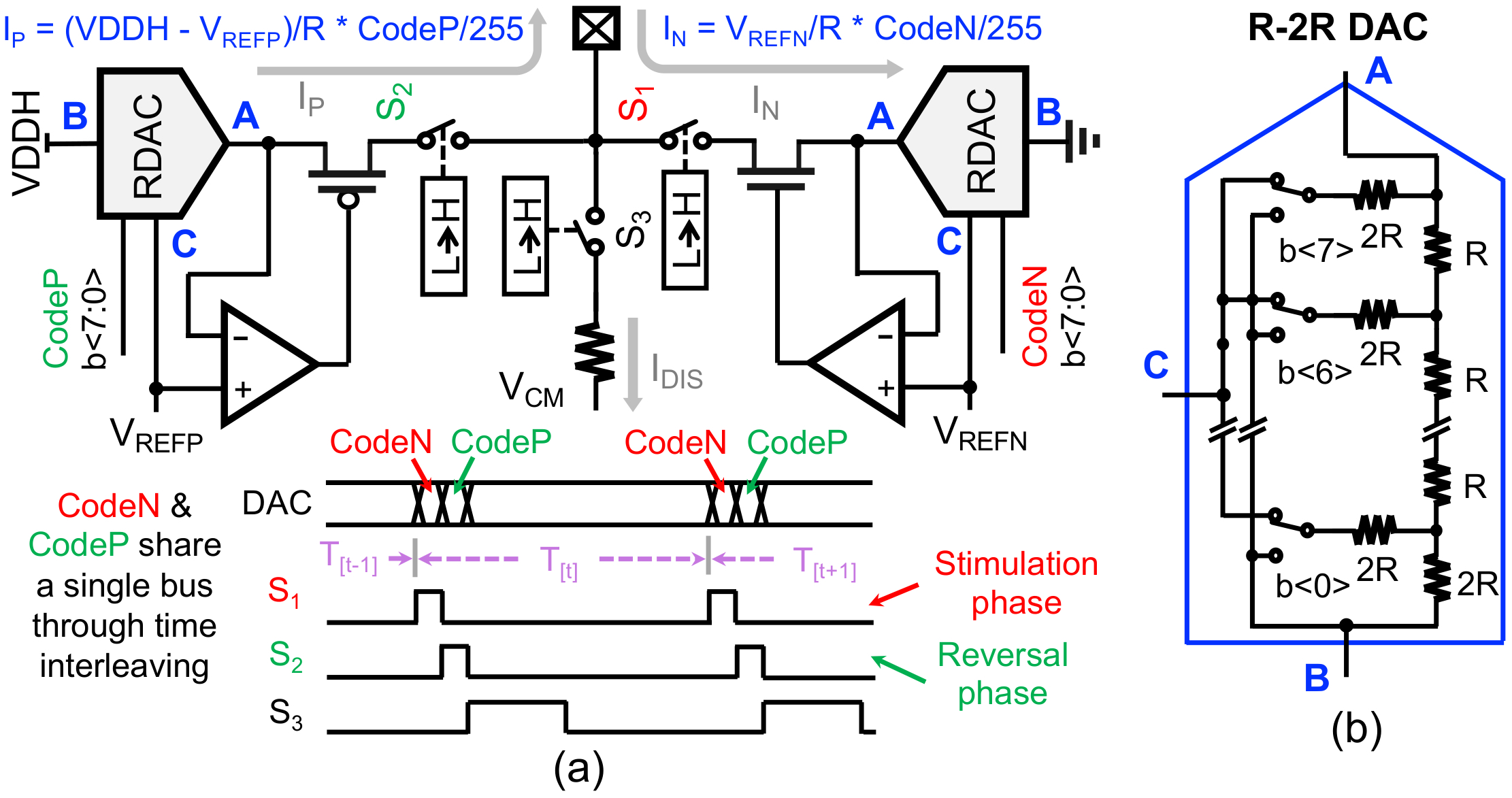}  \vspace{-3mm} \vspace{-1pt}
    \caption{(a) Schematic and timing of the current-regulated neural stimulator design. (b) The compact R-2R DAC used to set the currents.}
    \label{fig:stimulator}
\end{figure}

The firing-rate calculator (Fig.~\ref{fig:ckt_cl_controller}) accumulates each selected class in a cyclic bank of memory slices. Each slice stores events from one stride interval. At every stride boundary, the write pointer advances and the oldest slice is cleared and reused, while active slices are summed to produce $x$. Slice duration sets the stride and slice count sets the window length. Programmable coefficients select $y=\alpha x^2+\beta x+\gamma$ or $y=\alpha\log_2(x)+\gamma$. The quadratic mode covers linear, compressive, and expansive mappings. For $x=2^e(1+r)$, Mitchell's approximation gives $\log_2(x)\approx e+r$~\cite{mitchell1962computer}. A leading-one search extracts $e$, and normalized residual bits approximate $r$.
The result $y$ can control the stimulation interval, amplitude, and pulse width. For interval modulation, repeated accumulation of $y$ retains the phase residue and generates a carry that triggers the stimulation sequence (simplified timing example in Fig.~\ref{fig:ckt_cl_controller}). For amplitude and pulse-width modulation, $y$ directly sets the DAC codes and the durations of S1 and S2. This design enables flexible closed-loop mapping with minimal added hardware complexity.
Finally, a current-regulated stimulator (Fig.~\ref{fig:stimulator}) generates charge-balanced stimulation.
The current is set by two 8-bit R-2R DACs with a full-scale range adjustable via $\mathrm{V_{REFP}}$/$\mathrm{V_{REFN}}$ and set to 0.5~mA by default.

\vspace{-2mm}
\section{Experimental Results}

\begin{figure}[!h]
    \centering \vspace{-3mm}
    \includegraphics[width=.85\linewidth]{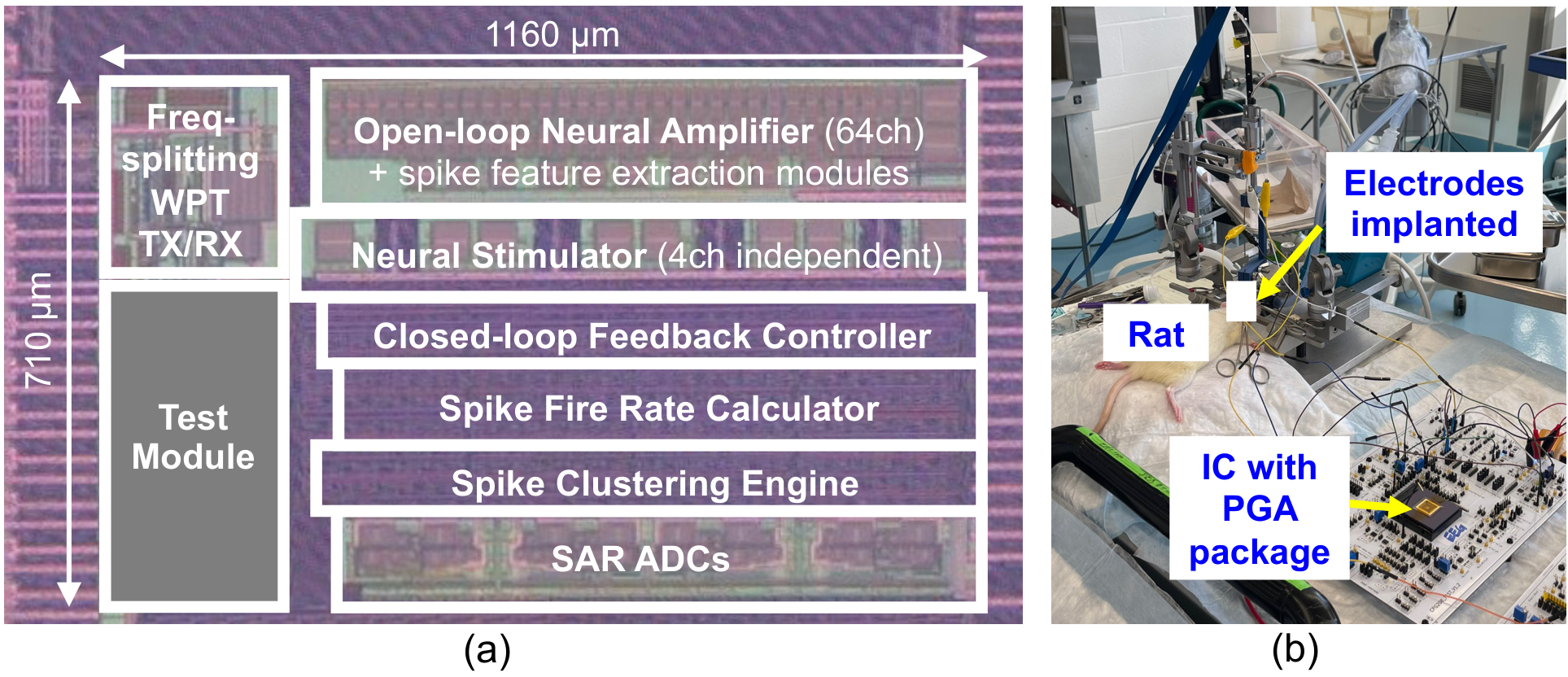} \vspace{-4mm}
    \caption{(a) A micrograph of the fabricated 65~nm chip with key blocks highlighted. (b) A photo showing an example \textit{in vivo} experimental setup.}
    \label{fig:die}
\end{figure}

The SoC has been fabricated in TSMC 65~nm GP. Fig.~\ref{fig:die}(a) shows the micrograph with key blocks highlighted. The chips were bench-tested and then validated \textit{in vivo} in rats (Fig.~\ref{fig:die}(b)).

\begin{figure}[!b]
    \centering \vspace{-3mm}
    \includegraphics[width=1\linewidth]{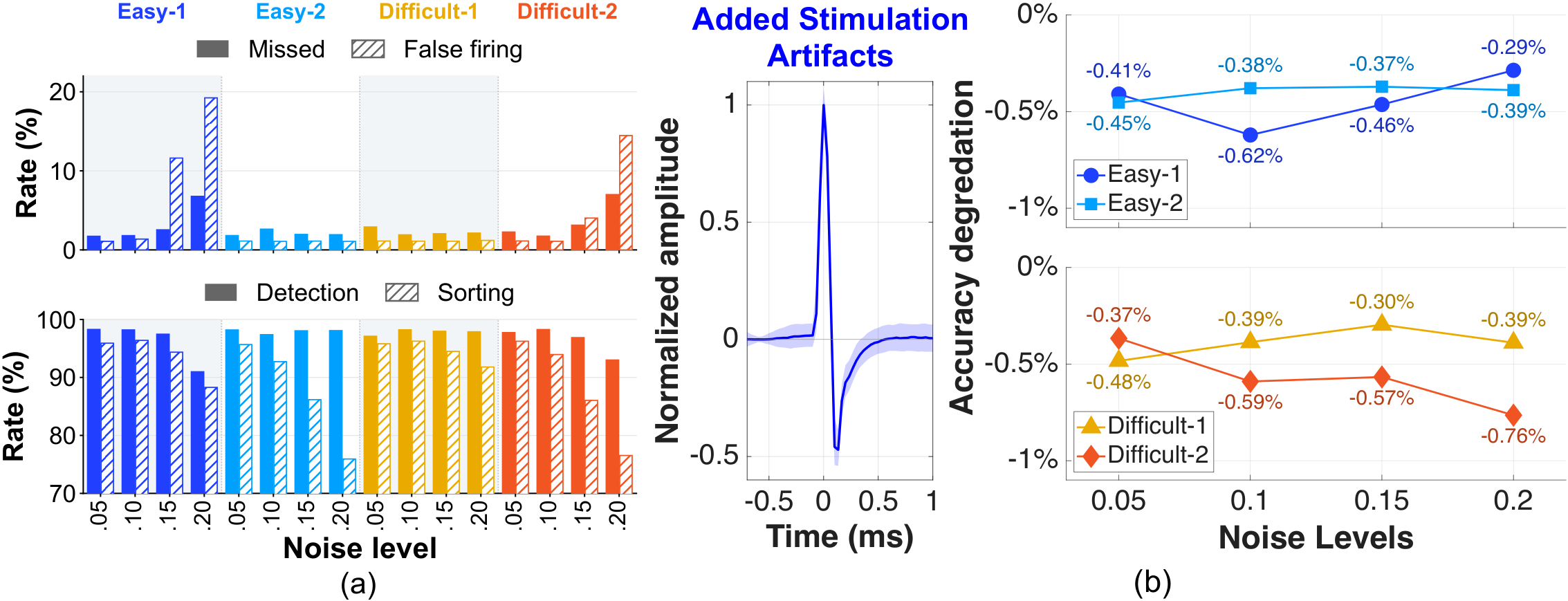} \vspace{-4mm} \vspace{-5pt}
    \caption{(a) Missed-spike and false-firing rates (top), and detection and sorting rates (bottom), across four datasets and noise levels. (b) Stimulation-artifact template derived from recorded artifacts and the resulting sorting accuracy degradation after artifact injection.}
    \label{fig:exp_sorting}
\end{figure}

The spike sorter was bench-tested by replaying four Quiroga benchmark sets, each at four noise levels~\cite{quiroga2004unsupervised}.
Signals were scaled below half the chip's input range.
The maximum cluster count $k$ is set to 3, $\eta$ to $1/32$, and $d_T$ to $1/16$ of full scale.
Fig.~\ref{fig:exp_sorting}(a) summarizes missed- and false-firing rates and detection and sorting accuracies. The average sorting accuracy is 91.03\%, reaching 94.58\% on Difficult-1, demonstrating competitive performance (Table~\ref{tab:spike}).
 Artifacts from our \textit{in vivo} recordings (Fig.~\ref{fig:exp_sorting}(b)) were scaled to amplitudes comparable to those of spikes and randomly injected at 20\% of the total spike count, avoiding exact spike overlap. The corresponding stimulation frequency averaged 11.6~Hz and peaked at 400~Hz.

\begin{table}[!t]
\centering
\caption{Comparison with State-of-the-Art Spike Sorters} \vspace{-2mm}
\label{tab:spike}
\scriptsize
\setlength{\tabcolsep}{1pt}
\begin{tabular}{!{\vrule width .4pt}M{.214\columnwidth}!{\vrule width .4pt}M{.132\columnwidth}!{\vrule width .4pt}M{.132\columnwidth}|M{.132\columnwidth}|M{.132\columnwidth}|M{.132\columnwidth}!{\vrule width .4pt}}
\Xhline{.4pt}
 & \tw\textbf{\makecell{This\\Work}} & \textbf{\makecell{ISSCC\\2023~\cite{chen2023spike}}}
   & \textbf{\makecell{JSSC\\2025~\cite{akhoundi2025scalable}}} & \textbf{\makecell{TBioCAS\\2025~\cite{hu2024processor}}}
   & \textbf{\makecell{JSSC\\2026~\cite{lukito2026spike}}} \\
\Xhline{.4pt}
\makecell{Implemen-\\tation} & \tw\makecell{Analog/\\Digital} & Digital & Digital & Digital & \makecell{Analog/\\Digital} \\
\hline
Power/ch & \tw 1.6~\uw & 1.78~\uw & 0.074~\uw & 2.08~\uw & 3.11~\uw \\
\hline
Dataset & \tw Quiroga & Quiroga & MEArec & Quiroga & Synthetic \\
\hline
\makecell{Reported\\Clust. Accuracy} & \tw 91.03\% & 89.50\% & \makecell{87.7\%/\\91.7\%} & 91.60\% & 94.54\% \\
\hline
\makecell{Accuracy (Diff. 1)} & \tw 94.58\% & 91.20\% & N/A & 94.60\% & N/A \\
\hline
Technology & \tw 65 nm & 22 nm & 40 nm & 65 nm & 65 nm \\
\hline
Stim. Art. Rej. & \tw Yes & N/A & N/A & N/A & N/A \\
\Xhline{.4pt}
\end{tabular} \vspace{-5mm}
\end{table}

\begin{figure}[!b]
    \centering \vspace{-5mm}
    \includegraphics[width=.9\linewidth]{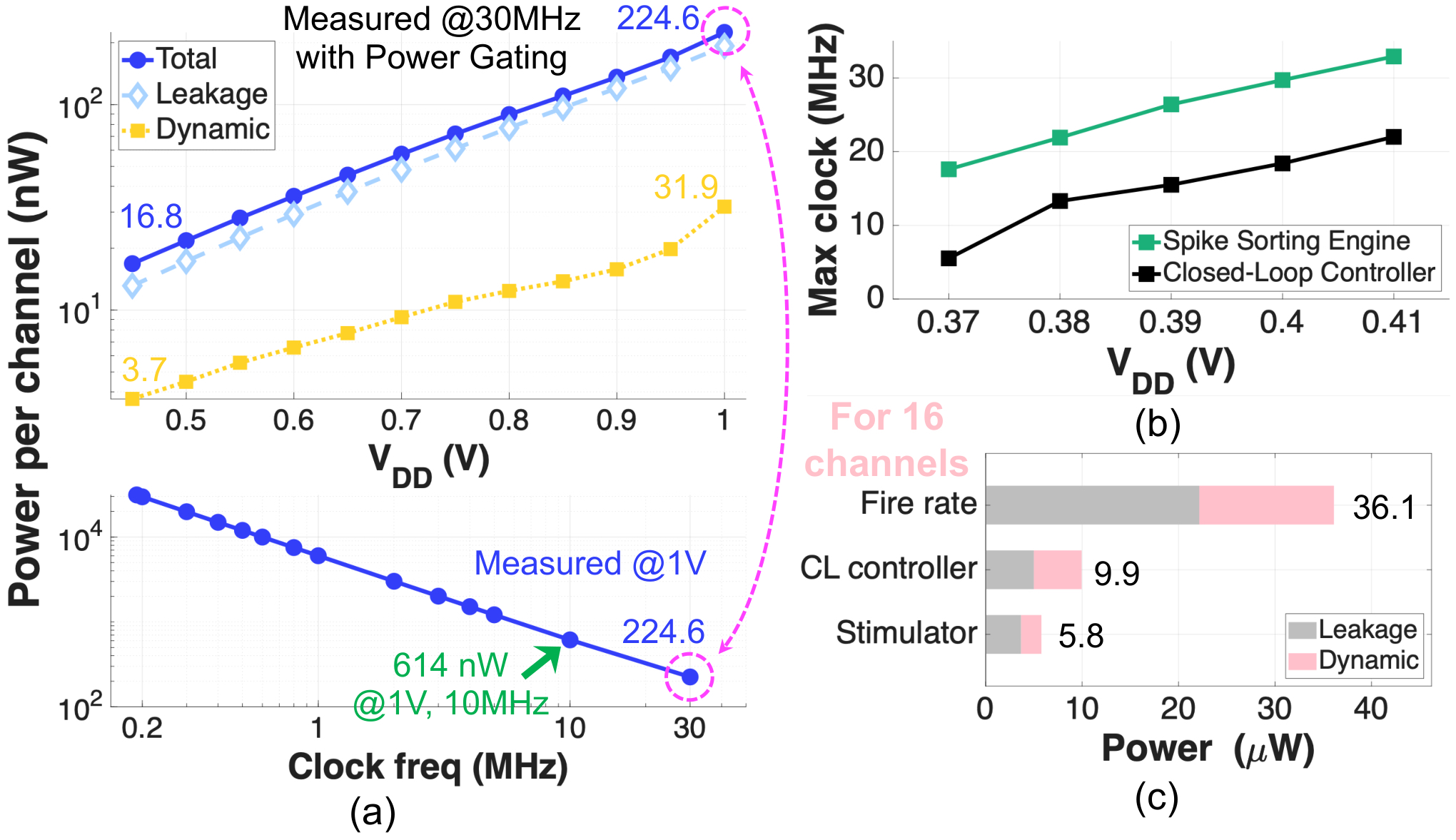} \vspace{-2mm} \vspace{-5pt}
    \caption{(a) Measured digital power of the spike sorter per channel with power gating. (b) Maximum clock frequency under different supply voltages. (c) Measured power of the main digital blocks.}
    \label{fig:exp_pwr}
\end{figure}

The complete spike-sorting path consumes 1.6\,$\mu$W/channel, including 0.98\,$\mu$W for the feature-extraction module (including the ADC) and 614~nW for the digital module at 1~V and 10~MHz with power gating, which is further decreased to 16.8~nW at 0.45~V and 30~MHz (Fig.~\ref{fig:exp_pwr}(a)). Since leakage dominates at low clock frequencies in this technology, faster processing reduces average digital power.
The maximum clock rates of the sorter and closed-loop controller are shown in Fig.~\ref{fig:exp_pwr}(b). The 16-channel rate calculator, controller, and stimulator logic consume 36.1\,$\mu$W, 9.9\,$\mu$W, and 5.8\,$\mu$W, respectively (Fig.~\ref{fig:exp_pwr}(c)).
Following firing-rate estimation (window dependent), the closed-loop controller's delay is under 2~$\mu$s.

\begin{figure*}[!t]
    \centering
    \includegraphics[width=\linewidth]{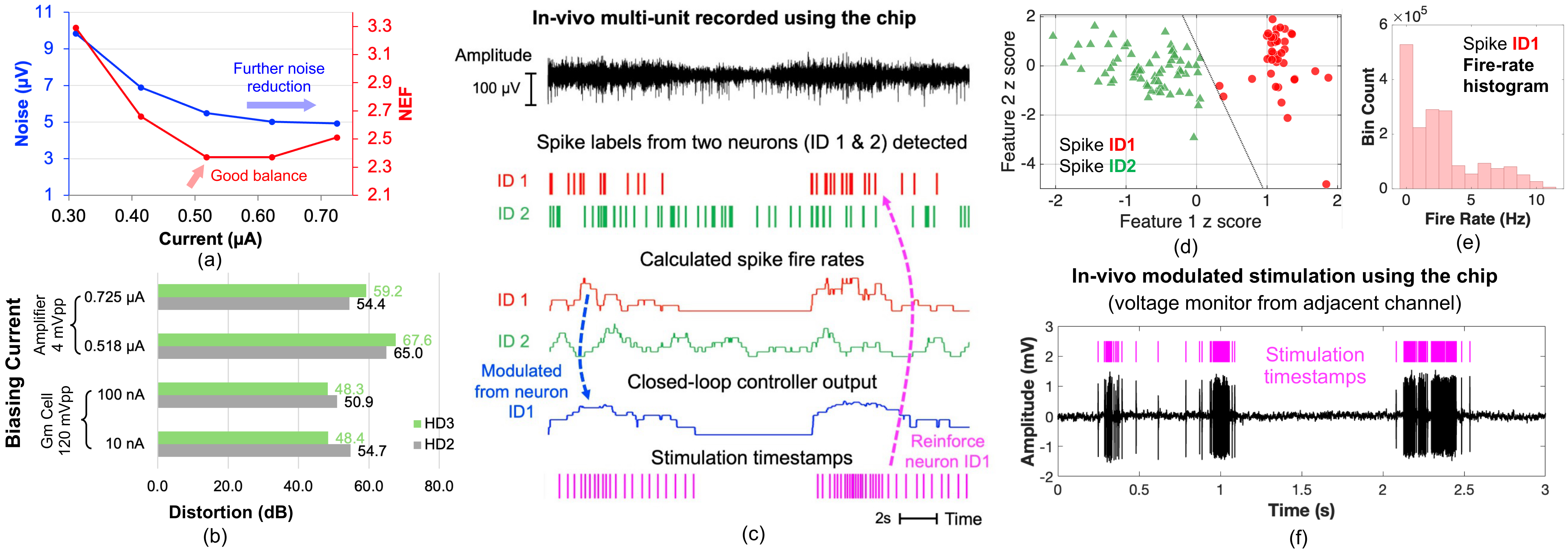} \vspace{-6mm} \vspace{-5pt}
    \caption{(a) Measured IRN and NEF of the OL amplifier showing the noise and efficiency trade-off. (b) Distortion of the OL amplifier and $g_m$ cell. (c) \textit{In vivo} recording and sorting of multi-unit activity and closed-loop mapping to stimulation, validating the feedback pathway. (d) Feature map showing well-separated neuronal clusters. (e) Firing-rate distribution matching expectation. (f) \textit{In vivo} modulated stimulation for functional verification and artifact characterization.} \vspace{-2mm}
    \label{fig:exp_wide}
\end{figure*}

Fig.~\ref{fig:exp_wide}(a) shows the measured input-referred noise (IRN) and noise-efficiency factor (NEF) of the OL amplifier from 300~Hz to 7~kHz with different biasing.
The selected bias achieves 5.5\,$\mu$V IRN and an NEF of 2.4, with HD2/HD3 below -65~dB for a 4~mV$_{\mathrm{pp}}$ input.
Further increasing power could reduce noise at the cost of efficiency.
The dynamic range is 55.2~dB with 1\% THD, the CMRR is 78~dB, and the input impedance is greater than 120~M$\Omega$. Monte Carlo simulations (100 runs) show a worst-case PSRR of 74.7~dB.
The measured $g_m$ cell achieves a wide tuning range with low distortion (Fig.~\ref{fig:exp_wide}(b)).
The IRN of the total analog front-end is 5.94\,$\mu$V, dominated by the OL amplifier.
Overall, analog feature extraction shows no statistical performance difference in sorting compared with the algorithmic implementation.

We conducted \textit{in vivo} experiments using the SoC in anesthetized rats.
Fig.~\ref{fig:exp_wide}(c) shows multiunit activity recorded and sorted into two classes, with their firing rates estimated independently and used for closed-loop stimulation mapping.
The feature map in Fig.~\ref{fig:exp_wide}(d) indicates well-separated neuronal clusters.
The firing-rate distribution of neuron ID~1 (Fig.~\ref{fig:exp_wide}(e)) is consistent with typical cortical activity, and this neuron is selected to modulate sensory stimulation.
The modulated stimulation was delivered to the animal and monitored through an adjacent channel (Fig.~\ref{fig:exp_wide}(f)) to verify timing and characterize stimulation artifacts at the front-end. These experiments validate the complete rapid-feedback signal path. Demonstrating neurophysiological benefits in behaving animals remains future work.
Additional chip specifications are summarized in Table~\ref{tab:summary} and compared against the state-of-the-art neuromodulation designs in Table~\ref{tab:comp}.

\begin{table}[!h]
\centering
\caption{Summary of Chip Specifications} \vspace{-1mm}
\label{tab:summary}
\scriptsize
\setlength{\tabcolsep}{2pt}
\renewcommand{\arraystretch}{1.0}
\begin{tabularx}{1\columnwidth}{|Y{1.0}|G{1.0}|Y{1.0}|G{1.0}|}
\hline
\textbf{Item} & \textbf{Value} & \textbf{Item} & \textbf{Value} \\
\hline
Chip tech. & 65 nm CMOS & Chip area & 0.82~mm$^2$ \\ \hline
OL Amp gain & 25\textendash30~dB & OL Amp BW & 300~Hz to 7~kHz \\ \hline
OL Amp noise & 5.5~$\mu$V$_\mathrm{rms}$ & OL Amp NEF & 2.4 \\ \hline
OL Amp DR & 55.2~dB (1\% THD) & OL Amp CMRR & 78~dB \\ \hline
ADC type & 11-bit SAR & ADC FoM$_\mathrm{S}$ & 156~dB \\ \hline
Sorter features & Max./min. slopes & Sorter distance & Manhattan \\ \hline
Sorter path power & 1.6\,$\mu$W/ch & Artifact loss & \resizebox{0.95\linewidth}{!}{$\approx$0.4\% at 20\% count} \\ \hline
CL controller ch. & 4 & CL controller delay & $<2~\mu\mathrm{s}$  \\ \hline
Feedback mapping & \resizebox{0.95\linewidth}{!}{2nd-order poly./$\log_2$} & Digital clock & 10~MHz \\ \hline
Stim. current & 8 bit, FS=0.5~mA & Stim. output & Charge-balanced \\ \hline
\end{tabularx}
\end{table}

\begin{table}[!h]
\centering \vspace{-2mm}
\caption{Comparison With State-of-the-Art Neural Interface SoCs} \vspace{-1mm}
\label{tab:comp}
\scriptsize
\setlength{\tabcolsep}{0.8pt}
\arrayrulecolor{black}
\begin{tabularx}{\columnwidth}{|Y{0.35}|Y{1.45}|G{1.05}|Y{1.05}|Y{1.05}|Y{1.05}|}
\hline
\multicolumn{2}{|c|}{\textbf{Publication}} &
\multicolumn{1}{c|}{\textbf{\makecell{This\\work}}} &
\textbf{\makecell{JSSC\\2026~\cite{zhang2026closedloop}}} & \textbf{\makecell{ISSCC\\2024~\cite{zhao2024interface}}} & \textbf{\makecell{ISSCC\\2024~\cite{hou2024multiloop}}} \\ \hline

\multirow{3}{*}{\rotatebox[origin=c]{90}{\textbf{General }  }}
 & Process & 65 nm & 40 nm & 180 nm & 180 nm \\ \cline{2-6}
 & Power rails & \makecell{0.45/1/\\2.5 V} & \makecell{0.73--1.1/\\1.8/3.3 V} & 1/1.25 V & \makecell{1.8/3.3/\\10 V} \\ \cline{2-6}
 & Application & \makecell{Prosthesis\\w/ feedback}
   & \makecell{Sleep\\Regulation} & \makecell{Neural\\Interface} & \makecell{Memory\\Enhancement} \\ \hline

\multirow{4}{*}{\rotatebox[origin=c]{90}{\makecell{\textbf{Neural}\\\textbf{Interface}}}}
 & Rec/Stim ch \# & 64/4 & 8/2 & 1$^*$ & 8 ch (x3) \\ \cline{2-6}
 & Amp Noise & 5.5 $\mu$V & 0.96 $\mu$V & 7 $\mu$V & 3.81 $\mu$V \\ \cline{2-6}
 & Amp NEF & 2.4 & - & 5.92$^*$ & 2.12 \\ \cline{2-6}
 & Stim mode & Current 8b & Optical & Optical & Current 8b \\ \hline

\multirow{3}{*}{\rotatebox[origin=c]{90}{\textbf{Closed-loop}}}
 & \makecell{Reported\\performance} & \makecell{94.58\%$^\dagger$\\(3-class)} & \makecell{82.4\%\\(5-class)} & - & - \\ \cline{2-6}
 & \makecell{Closed-loop\\algorithms} & \makecell{2nd order\\poly + log2} & DNN & - & EBM \\ \cline{2-6}
  & Reported Power & 1.6\,$\mu$W/ch & 0.2\,$\mu$J/class & - & 0.19\,$\mu$J/class \\ \hline

\end{tabularx}
\vspace{1pt}
\begin{flushleft} \vspace{-2mm}
\scriptsize $^*$Calculated from the reported data. $^\dagger$Measured on the Difficult-1 benchmark.
\end{flushleft}\vspace{-5mm}
\end{table}

\newpage

\section{Conclusion}
\vspace{-1mm}
This work demonstrates a 65~nm SoC that implements an on-chip feedback pathway from motor recording to sensory stimulation. Its event-driven mixed-signal feature extraction and adaptive Manhattan-distance-based spike sorting achieve high classification accuracy with robust stimulation-artifact rejection while consuming 1.6\,$\mu$W/channel. Sliding-window rate calculation and programmable polynomial or logarithmic mapping translate the activity of a selected neuron into stimulation parameters. Bench and \textit{in vivo} measurements validate the key recording, spike-sorting, firing-rate processing, mapping, and stimulation functions, establishing the feasibility of these techniques for future neuroprosthetic systems.

\bibliographystyle{IEEEtran}
\bibliography{ref}

@article{prsa2017rapid,
  title={Rapid integration of artificial sensory feedback during operant conditioning of motor cortex neurons},
  author={Prsa, Mario and Gali{\~n}anes, Gregorio L and Huber, Daniel},
  journal={Neuron},
  volume={93},
  number={4},
  pages={929--939.e6},
  year={2017},
  publisher={Elsevier},
  doi={10.1016/j.neuron.2017.01.023}
}

@article{goueytes2022learning,
  title={Learning in a closed-loop brain-machine interface with distributed optogenetic cortical feedback},
  author={Goueytes, Dorian and Lassagne, Henri and Shulz, Daniel E and Ego-Stengel, Val{\'e}rie and Estebanez, Luc},
  journal={Journal of Neural Engineering},
  volume={19},
  number={6},
  pages={066045},
  year={2022},
  publisher={IOP Publishing},
  doi={10.1088/1741-2552/acab87}
}

@INPROCEEDINGS{yu2025neuroprosthetic,
  author={Huang, Yu and Liu, Bowen and Hou, Yuhan and Xu, Jianxiong and You, Hao and Hung, Ashley and Ghosh, Swarnava and Liu, Eric and Yang, Naize and Ma, Junyu and Cai, Hanfeng and Kondrataviciute, Laura and Deng, Qiaosong and Kalia, Suneil K. and Richardson, Andrew G. and Hsieh, Ping-Hsuan and Genov, Roman and Liu, Xilin},
  booktitle={2025 IEEE International Solid-State Circuits Conference (ISSCC)}, 
  title={15.4 A Neuroprosthetic SoC with Sensory Feedback Featuring Frequency-Splitting-Based Wireless Power Transfer with 200Mb/s 0.67pJ/b Backscatter Data Uplink and Unsupervised Multi-Class Spike Sorting}, 
  year={2025},
  volume={68},
  number={},
  pages={272-274},
  doi={10.1109/ISSCC49661.2025.10904677}}

@article{mitchell1962computer,
  author  = {Mitchell, John N.},
  title   = {Computer Multiplication and Division Using Binary Logarithms},
  journal = {IRE Transactions on Electronic Computers},
  volume  = {EC-11},
  number  = {4},
  pages   = {512--517},
  month   = aug,
  year    = {1962},
  doi     = {10.1109/TEC.1962.5219391}
}

@article{quiroga2004unsupervised,
  author  = {Quiroga, Rodrigo Quian and Nadasdy, Zoltan and Ben-Shaul, Yoram},
  title   = {Unsupervised Spike Detection and Sorting with Wavelets and Superparamagnetic Clustering},
  journal = {Neural Computation},
  volume  = {16},
  number  = {8},
  pages   = {1661-1687},
  month   = aug,
  year    = {2004},
  doi     = {10.1162/089976604774201631}
}

@inproceedings{chen2023spike,
  author    = {Chen, Yingping and Tacca, Bernardo and Chen, Yunzhu and Biswas, Dwaipayan and Gielen, Georges G. E. and Catthoor, Francky and Verhelst, Marian and Mora Lopez, Carolina},
  title     = {A 384-Channel Online-Spike-Sorting IC Using Unsupervised Geo-OSort Clustering and Achieving 0.0013 mm$^2$/Ch and 1.78 $\mu$W/Ch},
  booktitle = {IEEE International Solid-State Circuits Conference},
  pages     = {486-487},
  month     = feb,
  year      = {2023},
  doi       = {10.1109/ISSCC42615.2023.10067264}
}

@article{hu2024processor,
  author  = {Hu, Zichen and Zhou, Zhining and Lyu, Hongming},
  title   = {A Power-and-Area-Efficient Channel-Interleaved Neural Signal Processor for Wireless Brain-Computer Interfaces with Unsupervised Spike Sorting},
  journal = {IEEE Transactions on Biomedical Circuits and Systems},
  volume  = {19},
  number  = {1},
  pages   = {108--119},
  month   = feb,
  year    = {2025},
  doi     = {10.1109/TBCAS.2024.3395353}
}

@inproceedings{zhao2024interface,
  author    = {Zhao, Linran and Shi, Wei and Gong, Yan and Liu, Xiang and Li, Wen and Jia, Yaoyao},
  title     = {A Miniature Neural Interface Implant with a 95\% Charging Efficiency Optical Stimulator and an 81.9 dB SNDR Delta-Sigma-Modulator-Based Recording Frontend},
  booktitle = {IEEE International Solid-State Circuits Conference},
  year      = {2024},
  doi       = {10.1109/ISSCC49657.2024.10454382}
}

@inproceedings{hou2024multiloop,
  author    = {Hou, Yuhan and Zhu, Yi and Wu, Xiao and Li, Yinfei and Lucas, Timothy and Richardson, Andrew G. and Liu, Xilin},
  title     = {A Multi-Loop Neuromodulation Chipset Network with Frequency-Interleaving Front-End and Explainable AI for Memory Studies in Freely Behaving Monkeys},
  booktitle = {IEEE International Solid-State Circuits Conference},
  year      = {2024},
  doi       = {10.1109/ISSCC49657.2024.10454413}
}

@article{zhang2026closedloop,
  author  = {Zhang, Chao and Zhang, Fenyan and Sheng, Dawid and Ma, Zhixiong and Guo, Yongxiang and Sun, Chao and Zhang, Yuwei and Zhao, Wenxin and Sheng, Xing and Wang, Tongfei A. and Zhang, Milin},
  title   = {An Electrophysiology-Optogenetics Closed-Loop Bi-Directional Neural Interface for Sleep Regulation With 0.2-$\mu$J/Class Multiplexer-Based Neural Network},
  journal = {IEEE Journal of Solid-State Circuits},
  volume  = {61},
  number  = {7},
  pages   = {3675--3687},
  month   = jul,
  year    = {2026},
  doi     = {10.1109/JSSC.2025.3639446}
}

@article{akhoundi2025scalable,
  author  = {Akhoundi, Arash and Yan, Pumiao and Landbrug, Yawende and Hays, Madeline and Murmann, Boris and Chichilnisky, E. J. and Muratore, Dante G.},
  title   = {A Scalable 1024-Channel Ultra-Low-Power Spike Sorting Chip With Event-Driven Detection and Spatial Clustering},
  journal = {IEEE Journal of Solid-State Circuits},
  volume  = {60},
  number  = {11},
  year    = {2025},
  doi     = {10.1109/JSSC.2025.3611139}
}

@article{lukito2026spike,
  author  = {Lukito, Vincent and Choi, Edward Jongyoon and Chang, Ik-Joon and Ha, Sohmyung and Je, Minkyu},
  title   = {A Spike Sorting {SoC} With $\Delta$-Based Spike Detection and End-to-End Implementation of Autoencoder Feature Extraction Using Analog {CIM}},
  journal = {IEEE Journal of Solid-State Circuits},
  volume  = {61},
  number  = {4},
  pages   = {1710--1721},
  month   = apr,
  year    = {2026},
  doi     = {10.1109/JSSC.2025.3589523}
}

\ifCLASSOPTIONcaptionsoff
  \newpage
\fi

\end{document}